# On the impact of smartification strategies for the state estimation of low voltage grids

Armin FATEMI[1], Franziska TISCHBEIN[1], Dr.-Ing. Frank WIRTZ[2],
Dr.-Ing. Christin SCHMOGER[3], Stefan DORENDORF[3],
Dr.-Ing. Annika SCHURTZ[4], Prof. Dr.-Ing. David ECHTERNACHT[4,5],
Univ. Prof. Dr. sc. Andreas ULBIG[1]

[1]IAEW RWTH Aachen, Schinkelstraße 6, 52062 Aachen, a.fatemi@iaew.rwth-aachen.de, f.tischbein@iaew.rwth-aachen.de, a.ulbig@iaew.rwth-aachen.de, www.iaew.rwth-aachen.de
[2]Bayernwerk Netz GmbH, Lilienthalstraße 7, 93049 Regensburg,
frank.wirtz@bayernwerk.de, www.bayernwerk-netz.de
[3]E.DIS AG, Langewahler Straße 60, 15517 Fürstenwalde/Spree,
christin.schmoger@e-dis.de, stefan.dorendorf@e-dis.de, www.e-dis.de
[4]E-Bridge Consulting GmbH, Baumschulallee 15, 53115 Bonn, aschurtz@e-bridge.de,
dechternacht@e-bridge.de, www.e-bridge.de
[5]Hochschule Düsseldorf, Münsterstraße 156, 40476 Düsseldorf,
david.echternacht@hs-duesseldorf.de, hs-duesseldorf.de

**Abstract:**

The decarbonization of for example the energy or heat sector leads to the transformation of distribution grids. The expansion of decentralized energy resources and the integration of new consumers due to sector coupling (e.g. heat pumps or electric vehicles) into low voltage grids increases the need for grid expansion and usage of flexibilities in the grid. A high observability of the current grid status is needed to perform these tasks efficiently and effectively. Therefore, there is a need to increase the observability of low voltage grids by installing measurement technologies (e.g. smart meters). Multiple different measurement technologies are available for low voltage grids which can vary in their benefit to observation quality and their installation costs. Therefore, Bayernwerk Netz GmbH and E.DIS AG in cooperation with E-Bridge Consulting GmbH and the Institute for High Voltage Equipment and Grids, Digitalization and Energy Economics (IAEW) investigated the effectiveness of different strategies for the smartification of low voltage grids. This paper presents the methodology used for the investigation and exemplary results focusing on the impact of intelligent cable distribution cabinets and smart meters on the quality of the state estimation.

# 1 Introduction

The decarbonization of for example the energy or heat sector leads to new challenges for distribution grids and their operators. The increasing amount of decentralized energy resources and the integration of new consumers due to competitiveness of electric vehicles and the transformation of the heat sector has the capability to bring the current distribution grids to their limits [1]. Many of these new units will be located in the low voltage (LV) level. To be able to tackle these challenges grid operators have to extend their grid effectively and efficiently. A possibility to reduce or shift the amount of future grid expansion needs can be active grid operation e.g. by means of controllable flexibilities. Both effective and efficient grid planning and active grid operation require observability of the current grid status. Many stakeholders in the energy sector have recognized this need for digitalization and smartification of the current systems [2]. For an accurate observability a reliable state estimation has to be executed which needs real measured values [3]. There are multiple technologies available for measurements in low voltage grids ranging from central measurements at the transformer substation to local measurements at the consumers utilizing smart meters. Therefore, there is the need to investigate which impact different equipment variants, using combinations of these technologies, can have on the state estimation in low voltage grids. This enables the grid operator to choose the most effective and cost-efficient strategy.

This paper presents a study which investigates the estimation quality of different equipment variants and relates them to the requirements of real-world use-cases like grid planning or active grid management. Therefore, we implement a method to calculate the state estimation quality for a given grid topology and equipment variant. The focus of this paper lies on the investigation of strategies based on intelligent cable distribution cabinets and smart meters.

# 2 Theoretical Background

This chapter explains the theoretical background used for the methodology and study we present. In a first subchapter, we introduce the mathematical background for state estimation and afterwards give an overview of the measurement technologies used in this study.

## 2.1 Mathematical Background

In grid operation most of the time not all information about the current grid status is available. Therefore, grid operators use a state estimation to calculate the missing values and the current grid status (e.g. voltage magnitudes and angles, power withdrawals and feed-ins). Thereby, the state estimation uses the current grid topology and available measurements. Using this information, the state estimation searches for a grid status which fits the given measurements the best minimizing the error squares. For our study we assume perfect knowledge about our current grid topology and only vary the given measurement locations (e.g. bus-voltage, power feed-in or current over a line) with the chosen equipment variant.

The state estimation algorithm builds a mathematical problem using the grid topology and the measurement values with their corresponding standard deviation $v$. The standard deviation describes the accuracy of the measurements which helps the algorithm to weight them amongst each other. The relationship of the state vector $\vec{x}$, representing the solution of the state estimation, and the measured values $\vec{z_m}$ is given by

$$\vec{z_m} = f(\vec{x}). \tag{1}$$

Because the measured values are fraught with uncertainty, the standard deviation $v$ is added in the mathematical model by

$$\vec{z_m} = \vec{z} + \vec{v}. \tag{2}$$

Therefore, the relationship described by (1) is replaced by

$$\vec{z_m} = f(\vec{x}) + \vec{v}. \tag{3}$$

Formula (3) has no analytical solution because the deviation vector $\vec{v}$ is unknown. It is possible to determine an estimated vector $\vec{\hat{x}}$ where the real measured values $\vec{z}$ deviate as little as possible from the given measured values $\vec{z_m}$ with the objective function

$$J = |z_m - z| \rightarrow min \tag{4}$$

which has its minimum at the position $\vec{\hat{x}}$. The described mathematical problem can then be solved by numerical methods.

## 2.2  Measurement Technology

For the equipment variants of this study, we include different technologies available for the measurement of low voltage grids. A summary about the technologies and their measurement location and values are displayed in Table 1. The first two technologies enable the observation of the medium-voltage (MV)-LV substations. Thereby, we differentiate between two types, digital Substations (digiONS) and intelligent substations (iONS). Digital substations are able to measure all electrical parameters in the substation. This includes the voltage at the busbars, the power and current which goes through the transformer and the power and currents which are exchanged with the underlaying grid through the feeders. Intelligent substations on the other hand are not able to measure the feeder of the substation, thereby missing the information how the power fed in to the grid is distributed on the different feeders. The third technology we include are intelligent cable distribution cabinets (iKVS). With iKVS the measurement of the bus and the lines connected to the conjunction box are possible. At last, we consider intelligent measurement technologies (iMSys), also called smart meters, as a possible measurement for consumers and generation units in low voltage grids. The focus of this paper is on iKVS and iMSys and how these technologies impact the state estimation of low voltage grids.

**Table 1:** Information about location and measured values of the measurement technologies

| Measurement technology | Measurement location | Measured value |
|---|---|---|
| digiONS | transformer | P, Q, I |
|  | busbar | V |
|  | feeder | P, Q, I |
| iONS | transformer | P, Q, I |
|  | busbar | V |
| iKVS | busbar | V |
|  | feeder | P, Q, I |
| iMSys | bus | P, Q, V, I |

As described in section 2.1 the state estimation considers the measured value in combination with their expected uncertainty. For the measurement technologies, investigated in this paper, the expected uncertainties are shown in Table 2. These uncertainties resemble the current technical standard available on the market. They correspond to the maximum tolerated measurement error. The measurement errors have a Gaussian distribution with a $3\sigma$ interval.

**Table 2:** Overview of the uncertainties for the measurement technologies

| Measurement | digiONS / iONS | iKVS | iMSys |
|---|---|---|---|
| Voltage (V) | 0.5 % | 0.5 % | 0.5 % |
| Current (I) | 1 % | 1 % | 1 % |
| Active Power (P) | 0.5 % | 0.5 % | 1 % |
| Reactive Power (Q) | 1 % | 1 % | 2 % |

## 3 Methodology

The goal of this study is to investigate the impact different smartification strategies for low voltage grids have on the state estimation quality. Therefore, we developed a method to vary the combination of measurement technologies and analyze how they affect the quality of the state estimation. The process of the method is shown in Figure 1.

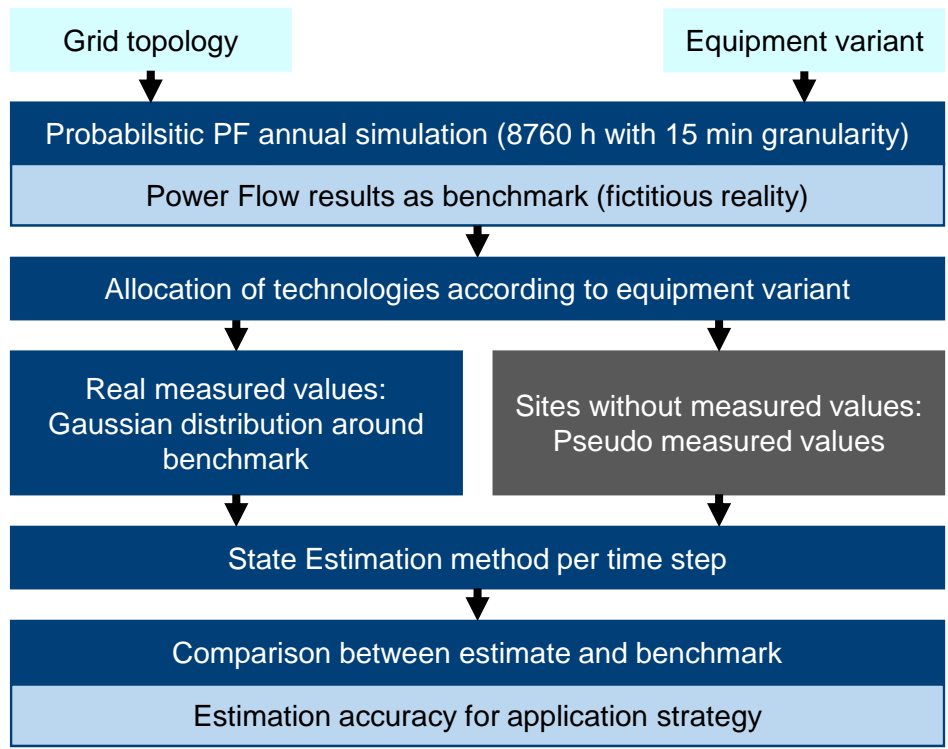

**Figure 1: Overview of the method of the study**

The inputs of the method are the grid topology and the equipment variant which is investigated. An equipment variant is a combination of the technologies described in 2.2, e.g. installed digiONS, 50 % iKVS and 10 % iMSys. The percentage is related to the maximum number of technologies that can be installed. The first step of the methodology is a power flow calculation on the given grid for the desired investigation timeframe, which is one year in this study with a granularity of 15 minutes. Therefore, annual power profiles are used for each consumer and generation unit in the grid. The results of the power flow function as a benchmark for comparison with the state estimation results to get the estimation quality for the equipment strategy. After the power flow calculation, the method allocates the measurement technologies according to the investigated equipment variant. The allocation methodology is further explained in 3.1. In the next step, the method generates measurement values for the technologies using the power flow results, further described in 3.2. Afterwards the method executes the state estimation for the timeframe using the generated measurement values. At the end the state estimation and power flow results can be used to calculate the desired evaluation parameters.

## 3.1 Allocation Method

In this section we describe the allocation method for iKVS and iMSys according to the investigated equipment variant.

### 3.1.1 Intelligent Distribution Cabinet (iKVS)

The allocation of the iKVS does impact the benefit they offer for the state estimation quality. Therefore, we chose the following priority for allocating the iKVS:

1. iKVS with the longest electrical distance to the transformer substation

2. iKVS with the highest number of feeders
3. random distribution

If the equipment variant requires an iKVS-distribution of 50 % we first allocate the iKVS with the longest electrical distance to the substation and afterwards the iKVS with the highest number of feeders. If the desired distribution of 50 % is still not met, we allocate the other iKVS randomly. Thereby, the iKVS with the longest electrical distance to the transformer substation provide significant information about the voltage of the substation feeder. The iKVS with the highest number of feeders carry valuable information about the power flows of the feeders.

### 3.1.2 Intelligent Measurement Technology (iMSys)

For the allocation of the iMSys we priorise the regulatory framework in Germany. Paragraph §29 of the Act on Metering Point Operation and Data Communication in Smart Energy Networks (MsbG) states that consumers with a consumption higher than 6.000 kWh per year are obligated to have an iMSys installed. Therefore, we chose the following priority:

1. Consumers with an annual consumption higher than 6.000 kWh
2. Consumers owning an electric vehicle
3. Random distribution

With this priorisation we are able to have a realistic allocation of our iMSys and therefore improve the practical usability of the investigation results.

## 3.2 Generation of Measurement Values

The state estimation needs a sufficient amount of measured values (2N-1) in the grid to be able to solve the mathematical problem. The quality of the state estimation depends on the number of measurement values that are available and the uncertainty these measurements have. In MV/LV grids the number of measurement technologies is scarce and therefore challenging the solvability of the problem. If additional information, e.g. the installed power and location of photovoltaic (PV) plants, exist we can support the state estimation by generating pseudo measured values. Therefore, we generate pseudo measured values for all consumers and generation units in the grid. In the following section we first describe the method to generate the measured values of the allocated technologies and afterwards the generation of the pseudo measured values.

### 3.2.1 Real measured values

The real measured values at the sites with measurement technology are generated by using the results from the power flow calculation which contains all the information using the annual power profiles of the consumers and generation units. To represent the uncertainties of the measurement technologies, introduced in Table 2, we draw a random number from a Gaussian distribution with the power flow result as the mean value and the standard deviation from the measurement technology.

### 3.2.2 Pseudo measured values

We add three different types of pseudo measured values to our state estimation. First, we generate pseudo measured values for the consumers in the grid. Therefore, the standard load profiles created by the BDEW can be used [4]. Different standard load profiles exist to represent different consumer groups (residential, agricultural, and commercial). We use the H0-profile for households, because they are the most common consumer in low voltage grids and scale them by the annual energy consumption, which we assume to be known by the grid operator.

In addition, we generate pseudo measured values for the PV-plants in the grid. Since low voltage grids span a small geographical area, there is an expected homogeneous weather pattern and therefore high correlation of the current-feed-in for all PV-plants in the grid [5]. Therefore, we use the annual profile of one of the PV-plants in the grid and scale it with the installed power of the other plants to generate their respective pseudo measured values. Thereby, we assume that grid operators know the location and installed power of the PV-plants in their grid.

For both consumers and PV-plants we assume a constant power factor of $\cos\Phi = 0.95$ and calculate their pseudo reactive power values via

$$Q = P \cdot \tan\Phi. \qquad (5)$$

At last we generate a pseudo measured value for the voltage at the point of common coupling with the overlaying medium-voltage (MV) grid using the result from the power flow calculation. Therefore, we assume that the MV-grids are completely observable by the grid operator.

## 4  Data

This chapter introduces the data used for the presented study. The data consists of the different investigated grid topologies and their corresponding supply task. In addition, we explain the different equipment variants we investigate in this study.

### 4.1  Grids and supply task

For this study we use four different low voltage grids from the Simbench-project [6]. An overview of the grid-describing data for all four grids is shown in Table 3. The target year for the supply task of the investigated grids is the year 2026 taking into account the grid development plan 2037/2045 [1].

**Table 3:** Overview of grid-describing data [6]

| Data | Rural 1 | Rural 2 | Semiurban | Urban |
|---|---|---|---|---|
| Feeders | 4 | 9 | 6 | 7 |
| Number of nodes | 96 | 128 | 110 | 58 |
| Number of cable distribution cabinets | 2 | 9 | 5 | 4 |
| Total line length | 1.47 km | 2.35 km | 1.79 km | 1.078 km |
| Number of consumers | 99 | 118 | 104 | 111 |
| Number of electric vehicles | 11 | 17 | 11 | 10 |
| Number of heat pumps | 8 | 18 | 14 | 14 |
| Number of PV-plants | 10 | 27 | 15 | 12 |
| Load/Generation (Max/Min) | 97 kW / -116 kW | 170 kW / -118 kW | 162 kW / 0 kW | 203 kW / -5 kW |

For the annual power flow calculation, the grid data is enriched with probabilistic profiles for all installed consuming and generating units [7] with a granularity of 15 min.

### 4.2 Equipment Variants

Our methodology introduced in chapter 3 allows the investigation of any combination of measurement technologies. An overview of all technologies and their technical details can be found in section 2.2. This study focuses on the impact of intelligent cable distribution cabinets and intelligent measurement technologies on the state estimation in low voltage grids. Therefore, we investigate the six variants, shown in Table 4, for all low voltage grids introduced in section 4.1. The first variant functions as a benchmark, where we have no measurement technologies in our grid and only use the pseudo measurements explained in 3.2.2. The second variant has only the digiONS installed to allow a general comparison of all technologies. Variant 3 and 4 represent a full rollout of 100 % iKVS and an assumed minimum rollout of 25 % to investigate the worst and best estimation quality achievable with iKVS. Variant 5 and 6 on the other hand focus on iMSys. First, we rollout a distribution of 11 % iMSys and afterwards 5 % iMSys. 11 % is an estimation of the number of consumers which will have an iMSys installed after the rollout in Germany within a low voltage grid. 5 % iMSys would be a worst-case assumption for grids which have less consumers obligated to install an iMSys by regulation. The allocation method of the technologies for each equipment variant and grid is explained in section 3.1.

**Table 4:** Overview of investigated equipment variants

| Variant | MV/LV-Substation | iKVS | iMSys |
|---|---|---|---|
| 1 | - | - | - |
| 2 | digiONS | - | - |
| 3 | - | 100 % | - |
| 4 | - | 25 % | - |
| 5 | - | - | 11 % |
| 6 | - | - | 5 % |

## 5 Results

In this chapter we present the results of our study focusing on the impact of iMSys and iKVS on the quality of the state estimation in low voltage grids. For all investigations we use the data (grid topologies, supply task and equipment variants) introduced in chapter 4.

### 5.1 Use-Cases and evaluation metric

In this section we first introduce the evaluation metric we base our results on. In addition, we define real-world use-cases for grid operators to be able to asses the relevance of our results.

Our evaluation metric consists of estimation quality for the voltages and currents in our grid. The estimation quality is the deviation of the state estimation in comparison with the real value, retrieved from the power flow calculation as described in chapter 3. For voltages the quality is defined by

$$E^{est}_{|V|} = \frac{|V_{est}| - |V_{pf}|}{|V_{pf}|} \qquad (6)$$

and for currents by

$$E^{est}_{|I|} = \frac{|I_{est}| - |I_{pf}|}{I_{th,max}} \qquad (7)$$

whereas $X_{est}$ is the result from the state estimation, $X_{pf}$ the result from the power flow and $I_{th,max}$ the maximum thermal current for the corresponding line. Therefore, $E^{est}_{|I|}$ is a measure for how much of the line loading cannot be used due to the estimation error. In the following investigations we will calculate these evaluation parameters for each element and one year. Due to the granularity of 15 min, in total 35040 values result for each bus and line. For each grid and equipment variant we then consider the 99 % quantile for the voltage magnitudes and the 95 % quantile for the line magnitudes over all quality values [5]. This allows us to compare the performance of different equipment variants in the different low voltage grids.

For distribution grid operators there are multiple use-cases for state estimation in low voltage grids. For this study we define three different use-cases based on preliminary analysis: grid planning, connection request and monitoring / active grid management. In addition, we set a requirement for the estimation quality for each use-case, presented in Table 5. Grid planning can use the state estimation results to determine the optimal time for expansion measures more precisely. For connection requests, e.g. for new PV-plants, the state estimation can improve the information about the current grid status and therefore optimize the decision making and utilization of current grid assets. Higher requirements on specially line loading are needed because the decision in the connection request could lead to possible congestions if based on wrong information. Monitoring of the current grid during live-operation (in "real-time") is possible with the results from the state estimation. Based on the monitoring, active grid operation can be executed to optimize the current grid utilization by for example executing swichting measures. Active grid operation has a high requirement on the state estimation quality because the resulting actions have immediate effect on the grid operation.

**Table 5:** Overview of estimation requirements of grid operator use-cases

| Use-Case / Estimation parameter | Grid Planning | Connection Request | Monitoring / Active Grid Management |
|---|---|---|---|
| Voltage magnitude | ± 2 % | ± 1.5 % | ± 1 % |
| Line loading | ± 10 % | ± 5 % | ± 5 % |

## 5.2  General comparison of measurement technologies

In our first investigation we aim to get an overview of the performance of the different measurement technologies using variant 1, 2, 3 and 5 from Table 4. Figure 2 shows the comparison of the 99 % quantile for the voltage magnitudes for the chosen equipment variants and grids. The reference variant where no measurement technology is installed is mostly not able to meet the requirements for all three use-cases. The best-case variants for the other technologies are all able to meet the requirements of all use-cases. The differences for the performance between the grids are also insignificant.

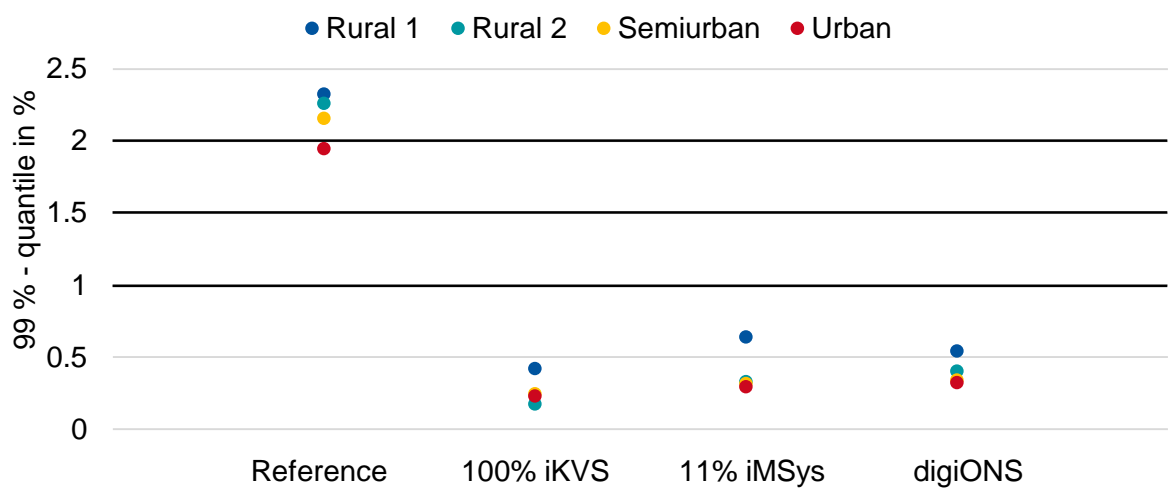

**Figure 2:** Estimation quality for the voltage magnitudes of the general comparison of measurement technologies

Figure 3 shows the 95 % quantile for the line loadings. In this case the reference variant is able to meet the requirement for grid planning reaching a quality below 10 % deviation for all grids. The requirements of the other two use-cases cannot be fulfilled for the reference case. The digiONS and a distribution of 100 % iKVS are able to fulfill all requirements reaching qualities between 2 % and 4 %. 11 % iMSys are not able to reach a quality below 5 % for most grids. Therefore, doing active grid management and monitoring in real-time only with 11 % iMSys is not advised. Additional technology should be added, like a combination of digiONS and 11 % iMSys.

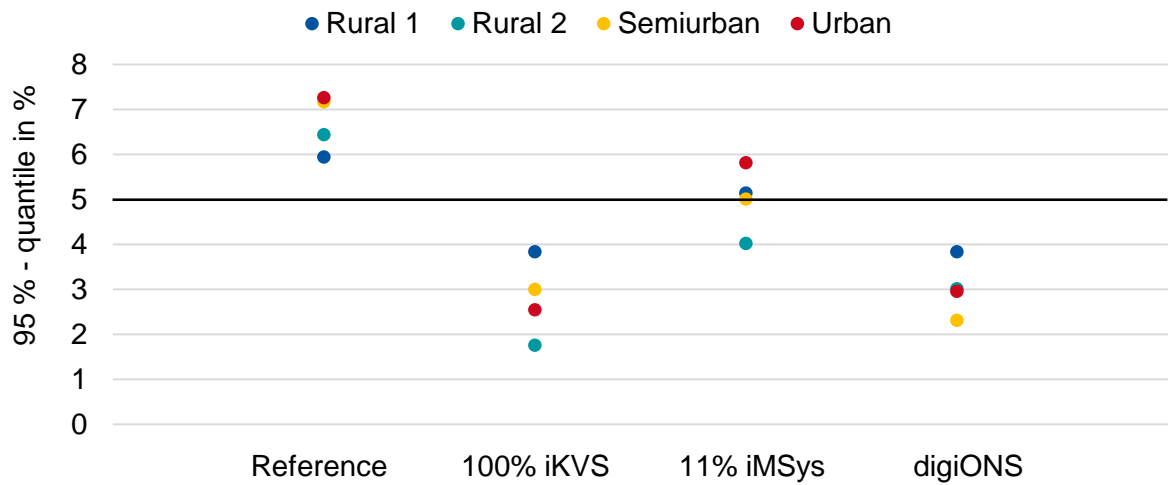

**Figure 3:** Estimation quality for the line loadings of the general comparison of measurement technologies

## 5.3   Comparison of iKVS strategies

In our second investigation we want to analyze if a strategy following the rollout of iKVS would be able to meet the requirements of all use-cases if only 25 % iKVS would be installed. The comparison with 100 % iKVS will get us the best- and worst-case view. Figure 4 shows the 99 % quantile for the voltage magnitudes of the two variants for all grids. The variant with 25 % iKVS is also able to meet all the requirements reaching estimation qualities below 0.7 %. Comparing the qualities of the both variants we can see that the performance is spread between the grids for the 25 % variant, while 100 % iKVS has rural 1 as an outlier performing worse than the other three grids. This can be explained by looking at the total number of possible iKVS from the data in Table 3. Because in rural 1 only two iKVS can be installed the quality increase is less than the other grids.

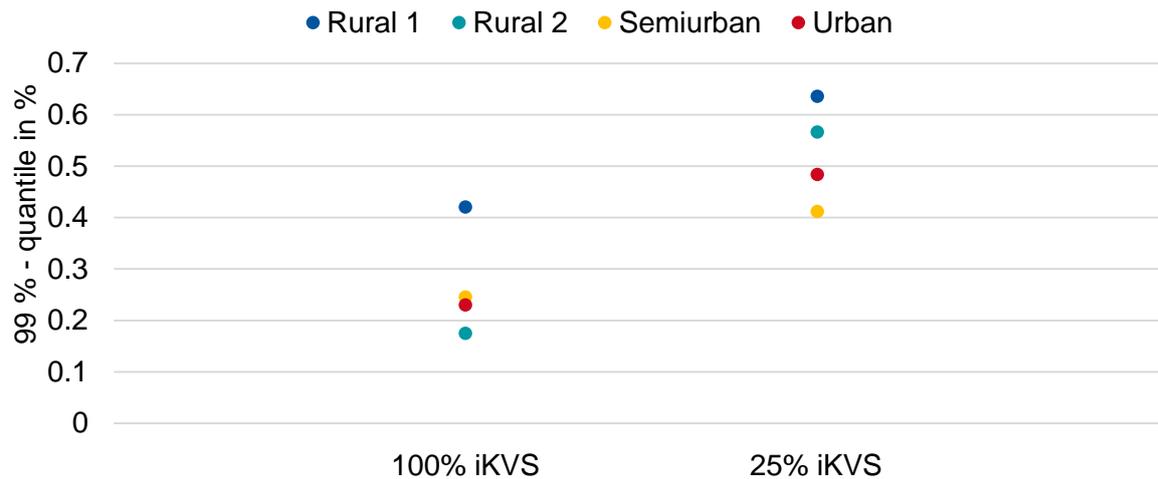

**Figure 4:** Estimation quality for the voltage magnitudes of the comparison of iKVS distributions

Figure 5 shows the 95 % quantile for the line loading estimations for the two variants. In this case, decreasing the distribution from 100 % to 25 % iKVS leads to a surpass of our limit for the use-cases connection request and monitoring / active grid management. With 25 % iKVS there are feeders in the grids without any measurement technology installed which leads to the state estimation having worse information what power is going into or from the feeder. This results in a decrease of estimation quality. Therefore 25 % iKVS is not suitable for these tasks similar to 11 % iMSys. Because the qualities are very close to the requirement of 5 % an increase of the distribution would likely resolve the issue. In addition, a combination with other technologies like the digiONS would be a possible solution.

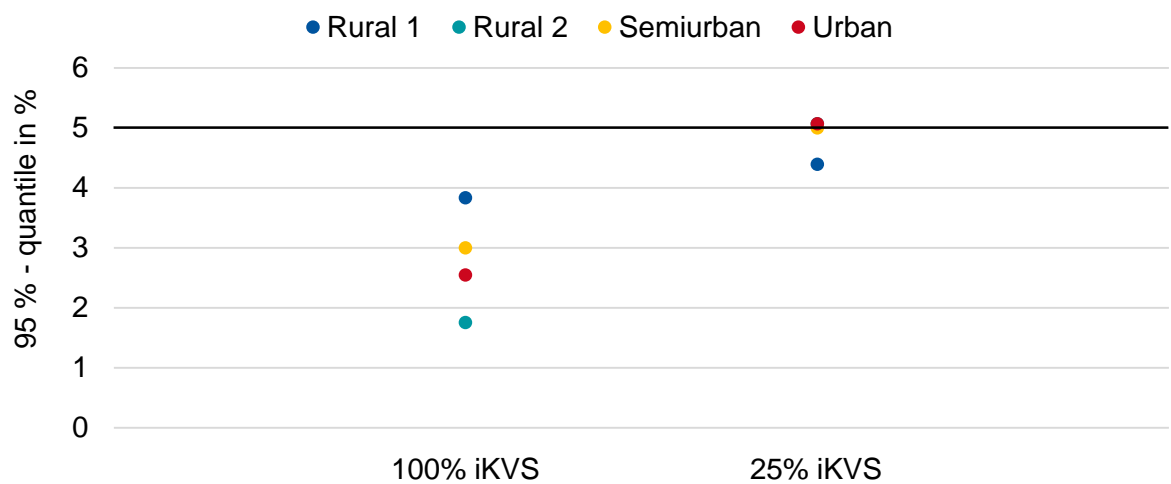

**Figure 5:** Estimation quality for the line loadings of the comparison of iKVS distributions

Similar to the voltages rural 1 performs worse than the other three grids for 100 % iKVS. For 25 % iKVS on the other hand it shows the best performance of all grids. This can be explained by the location of the iKVS in relation to the consumers in the grid. Figure 6 shows the location of PV-plants, electric vehicles, consumers with a yearly consumption higher than 6.000 kWh and consumers not fitting the other three categories in the topology of rural 1. In addition, the location of the iKVS is marked according to the allocation method, described in 3.1, which chooses the iKVS farthest away from the transformer. Most big consumers in the grid are in one feeder which is also the one the iKVS is placed in. This leads to a significant rise in the estimation quality. For the other grids the consumers and iKVS are more spread. Overall, we can observe that the performance of iKVS depends on the location of the iKVS. iKVS placed in long feeders or feeders with high amount of load benefit the state estimation the most. Therefore, the allocation of iKVS should be decided based on the individual grid.

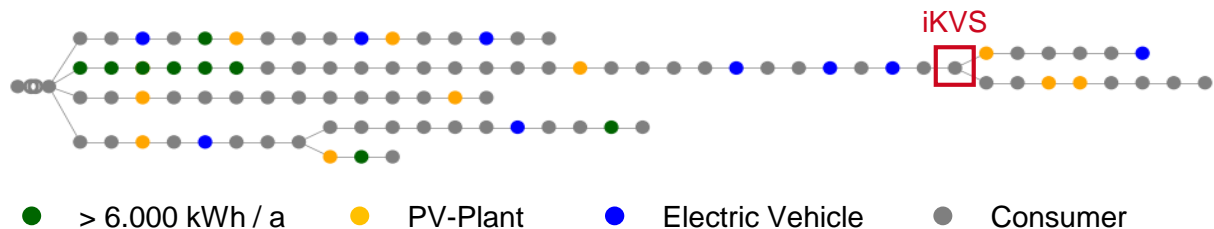

**Figure 6:** Overview of the location of consumer types and allocated iKVS for rural 1

### 5.4 Comparison of iMSys strategies

Our last investigation focuses on the rollout of intelligent measurement technologies. Therefore, we want to investigate how a worst-case rollout of 5 % iMSys would impact the quality of the state estimation. Figure 7 shows the 99 % quantile for the voltage magnitudes of all four grids with 11 % iMSys and 5 % iMSys. 5 % iMSys can still satisfy all requirements of our use-cases. As in our second investigation rural 1 is an outlier again performing worse than the others grids for both variants. The reason is once again the location of the consumers, shown in Figure 6. Based on regulation our allocation method first allocates iMSys to the consumers with a yearly consumption over 6.000 kWh, as described in 3.1. Because all these consumers are mostly in one feeder in rural 1 we have less information about the overall grid. For the other three grids the technologies are more spread, giving more information about the other feeders. Therefore, as with iKVS, the placement of iMSys is crucial for the quality of the state estimation.

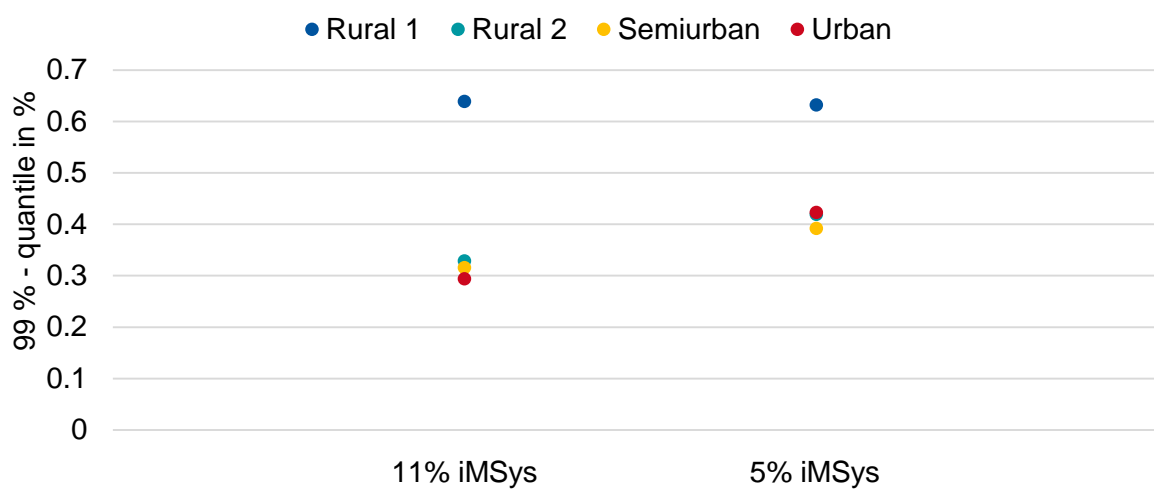

**Figure 7:** Estimation quality for the voltage magnitudes of the comparison of iMSys distributions

Figure 8 shows the 95 % quantile for the line loadings of all grids. Because 11 % could not reach the requirement of all use-cases for all grids, we added the variant of 5 % iMSys with the digiONS. This allows us to not only compare the loss in quality by only having 5 % iMSys but also evaluate if the combination with digiONS can still reach the requirements. As expected, 5 % iMSys alone reduces the estimation quality compared to 11 % iMSys, now surpassing the 5 % quality limit for all grids. The combination with digiONS creates a significant improvement being able to satisfy all requirements.

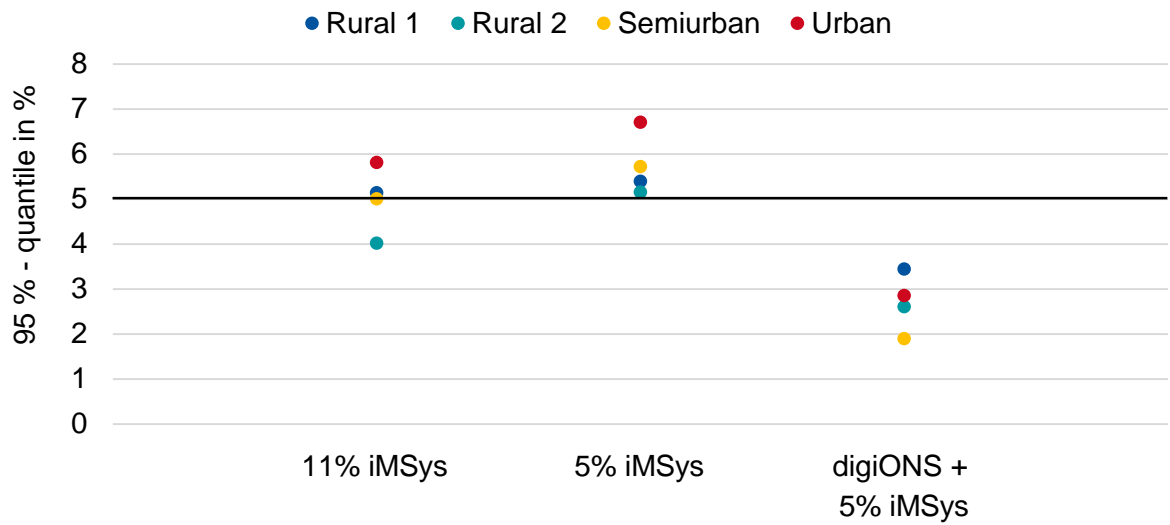

**Figure 8:** Estimation quality for the line loadings of the comparison of iMSys distributions

## 6   Summary and Outlook

In this paper we presented a study for the quality of state estimation in low voltage grids based on different type of equipment variants with measurement technologies. The presented methodology is able to allocate any combination of measurements for transformer substations, cable distribution cabinets and smart meters to a given grid. Based on the chosen equipment variant, e.g. 50 % intelligent cable distribution cabinets (iKVS) and 10 % intelligent measurement technologies (iMSys), a state estimation is calculated for one year on the given grid topology. For this study the focus lies on the impact of iKVS and iMSys on the state estimation quality. Therefore, requirements for the grid operator use-cases: grid planning, connection request and monitoring / active grid management where analyzed and defined beforehand.

In summary the results show the following:

- The quality of the state estimation is highly dependent on the location of iKVS and iMSys. A placement on feeders with big consumers or at the end of feeders are preferable.
- 100 % iKVS are able to reach all requirements of the use-cases while 25 % iKVS are not.
- 11 % iMSys and 5 % iMSys are both not able to satisfy all requirements. In both cases a combination with for example a measurement at the transformer substation can resolve the issue.

In future work we will verify these results by investigating further grid topologies and supply tasks. In addition, the combination of iKVS and iMSys should be investigated and also what impact uncertainty of the grid topology data would have on the estimation quality. Another aspect which has to be investigated in future work are the costs of the equipment variants. For the strategy of a grid operator the estimation quality in relation to the resulting costs are the relevant parameter. Therefore, a cost analysis for each technology has to be done so a comparison of the variants is possible.

# 7 References

[1] Bundesnetzagentur, "Genehmigung des Szenariorahmens 2023-2037/2045," Bonn, 2022.

[2] European Commission, "Actions to digitalise the energy sector", 2022, [Online], Available: https://ec.europa.eu/commission/presscorner/detail/en/ip_22_6228

[3] D. Echternacht *et al.,* "Smart Area Aachen - in field test of meter placement and state estimation algorithms for distribution grids," in *2015 IEEE PES Innovative Smart Grid Technologies Latin America (ISGT LATAM)*, 2015, pp. 435–439.

[4] VDEW, "Repräsentative VDEW-Lastprofile," Frankfurt (Main), 1999.

[5] David Echternacht, *Optimierte Positionierung von Messtechnik zur Zustandsschätzung in Verteilnetzen*. Aachen, 2015.

[6] Steffen Meinecke *et al.,* "SimBench--A Benchmark Dataset of Electric Power Systems to Compare Innovative Solutions based on Power Flow Analysis," *Energies*, vol. 13, no. 12, p. 3290, 2020, doi: 10.3390/en13123290.

[7] C. M. Vertgewall, C. Holscher, L. R. Böttcher, et.al., "Modeling and Application of Probabilistic Electrical Household Loads in Distribution Grid Simulations" in *2022 International Conference on Smart Energy Systems and Technologies (SEST)*, Eindhoven, Netherlands, 2022, pp. 1-6, doi: 10.1109/SEST53650.2022.9898438.